\documentstyle[12pt]{article}

\newcommand{\la}[1]{\label{#1}}
\newcommand{\beq}{\begin{equation}}
\newcommand{\eeq}{\end{equation}}
\newcommand{\ba}{\begin{eqnarray}}
\newcommand{\ea}{\end{eqnarray}}
\newcommand{\bea}{\begin{eqnarray*}}
\newcommand{\eea}{\end{eqnarray*}}
\let\oref=\ref
\renewcommand{\ref}[1]{(\oref{#1})}
\newcommand{\del}{\partial}

\makeatletter
\def\fnum@figure{Fig.~\thefigure}
\newdimen\ctcleftskip \ctcleftskip=.4em
\newdimen\ctcd@pth \newdimen\ctcsk@p \ctcsk@p=0pt
\def\c@ntract#1#2{\ctcd@pth#2\relax\mathord{\vtop{\ialign{##\crcr
$\displaystyle{#1}$\crcr\noalign{\kern1pt\nointerlineskip}
\ctcf@ll\crcr}}}}
\def\ctcf@ll{\hskip\ctcsk@p\hskip\ctcleftskip\vrule\@height\ctcd@pth
\@depth\z@\leaders\hrule\hfill%
\vrule\@height\ctcd@pth\@depth\z@\hskip\ctcleftskip\relax}
\newdimen\ctcnormaldepth \ctcnormaldepth=2.5pt
\def\contract#1{\c@ntract{#1}{\ctcnormaldepth}}
\def\xcontract#1#2{\c@ntract{#2}{#1\ctcnormaldepth}}
\def\doublecontract#1#2#3{
\setbox\z@\hbox{$\displaystyle\contract{#1#2}$}
\setbox\@ne\hbox{$\displaystyle#1$}\dp\z@\dp\@ne\ctcsk@p=\wd1
\setbox\@ne\hbox{$\displaystyle\xcontract2{\box\z@#3}$}
\box1\ctcsk@p\z@}
\def\innercontract#1{
\setbox\z@\hbox{$\displaystyle\contract{#1}$}
\setbox\@ne\hbox{$\displaystyle #1$}\dp0=\dp1\box\z@}
\makeatother

\begin{document}
\pagestyle{empty}
\setlength{\parindent}{0.0cm}

\begin{center}
\vspace*{2.2cm}
{\Large\bf Gauge Dependence of the High-Temperature} \\
\vspace*{2mm}
{\Large\bf 2-Loop Effective Potential for the Higgs Field} \\
\vspace*{12mm}
{\large\bf M.~Laine\footnote{
Email: mlaine@phcu.helsinki.fi}} \\
\vspace*{3mm}
{\sl  Department of Theoretical Physics, \\
P.O.~Box 9, FIN-00014 University of Helsinki, Finland} \\
\vspace*{3mm}
9 November 1994
\end{center}

\vspace*{-9.8cm}
\hfill Preprint HU-TFT-94-46
\vspace*{9.5cm}

\begin{center}
{\large\bf Abstract}
\end{center}
\vspace*{2mm}
The high-temperature limit of the 2-loop effective potential for
the Higgs field is calculated from an effective 3d theory, in a
general covariant gauge. It is shown explicitly that a gauge-independent
result can be extracted for the equation of state from the gauge-dependent
effective potential. The convergence of perturbation theory is estimated
in the broken phase, utilizing the gauge dependence of
the effective potential.

\newpage
\pagestyle{plain}
\setcounter{page}{1}
\setcounter{footnote}{0}
\setlength{\parindent}{0.85cm}

\section{Introduction}

Recently, the high-temperature electroweak phase transition
has been the subject of active research, due to its
possible effect on the baryon number of the Universe.
The standard tool for perturbative investigations of
the equilibrium properties of this phase transition
is the effective potential for the Higgs field, $V(\varphi)$.
{}From the effective potential, one can calculate quantities like
the order, the critical temperature, and the latent heat
of the phase transition. In particular, the pressure is given
by minus the value of the effective potential at the minimum.
Presently, the most complete
expressions for~$V(\varphi)$ are given by the
2-loop calculations in refs.~\cite{ArEs,FoHe,FKRS1}.


When calculating the effective potential,
one of the problems one is faced with is gauge dependence.
Indeed, the source term appearing in the generating functional
\beq
\exp({-W[J]})\equiv
\int{\cal D}\phi{\cal D}A\exp({ -S[\phi,A]-
(J^{\dagger}\Phi+\Phi^{\dagger}J)/2})
\la{source}
\eeq
is not gauge-invariant, and hence $W[J]$ may be gauge-dependent.
The effective potential~$V(\varphi)$, which is obtained from the Legendre
transformation of $W[J]$, then also
depends on the gauge condition~\cite{Ja}.
In particular, the location of the minimum of the effective
potential is gauge-dependent, which
indicates that $\varphi$ is not a physical observable.
However, the value of the effective potential at a
stationary point is obtained by putting
the source $J$ to zero, and
consequently one should obtain a gauge-independent result
for the pressure~\cite{Ni,FuKu}.
It is the main purpose of this paper to check explicitly
to 2-loop order~($\hbar^2$) that a gauge-independent result
is, indeed, obtained.

There are methods of extending the definition of the effective potential
so that it is gauge-independent even away from the
minima~(see, e.g.,\ ref.~\cite{Vi}).
A concrete example,
relevant for the EW phase transition, was given in ref.~\cite{BFH}.
There the external source was coupled to the composite
operator $\Phi^{\dagger}\Phi$:
\beq
\exp({-\widetilde{W}[J]})\equiv
\int{\cal D}\phi{\cal D}A\exp({ -S[\phi,A]-2J\Phi^{\dagger}\Phi})
\,\, .
\la{desy}
\eeq
The source term $2J\Phi^{\dagger}\Phi$, and hence both $\widetilde{W}[J]$
and the corresponding effective potential~$\widetilde{V}(\sigma)$, are
manifestly gauge-independent. The pressure is again obtained
from the value of~$\widetilde{V}(\sigma)$ at its minimum,
or, equivalently, from~$\widetilde{W}[0]$. It is checked
below to order~$\hbar^2$ that $\widetilde{V}(\sigma)$ and
the conventional gauge-dependent effective potential $V(\varphi)$ give
the same result for the pressure.

The paper is organized as follows.
In Sec.~\oref{details}, some technical details of the
calculation of the 2-loop effective potential in
a general covariant gauge are discussed.
In Sec.~\oref{eos}, it is shown that the value of the effective
potential at the minimum is gauge-independent to order $\hbar^2$.
In Sec.~\oref{giep}, it is shown that the effective
potentials resulting from eqs.~\ref{source} and~\ref{desy} yield
the same values for the pressure, when calculated consistently
in powers of $\hbar$.
The calculation of some other physical quantities than pressure
is discussed in Sec.~\oref{physical}.
In Sec.~\oref{con}, the convergence of perturbation theory
is studied in the broken phase,
utilizing the gauge-dependent effective potential.
The conclusions are in Sec.~\oref{clu}.
The explicit form of the 2-loop effective potential is
represented in the Appendix. Sections~\oref{details}
and~\oref{con} are an extension of ref.~\cite{La}.

\section{Method of calculation}
\la{details}

In the high-temperature limit, the essential long-wavelength
properties of the Standard Model can be described by an effective
super-renormalizable 3-dimensional field theory~\cite{ApPi,Na}.
This dimensional reduction is accomplished by integrating
out all the degrees of freedom corresponding to the momentum
scale $p\sim T$ in the original 4d theory.
At the 2-loop level, it has been explicitly verified~\cite{FKRS1}
that the 3d theory produces the high-temperature limit
of the 4d effective potential for the Higgs field.
Furthermore, the parameters of the effective 3d theory
should be independent of the 4d gauge fixing condition~\cite{FKRS1,JaPa}.
The relevant 3d action is
\ba
S\hspace*{-2mm} & = &\hspace*{-2mm}
\int\! d^3x \biggl[
\frac{1}{4}F^a_{ij}F^a_{ij}+ (D_i\Phi)^{\dagger}(D_i\Phi)+
m_3^2\Phi^{\dagger}\Phi+\lambda_3
(\Phi^{\dagger}\Phi)^2+ h_3\Phi^{\dagger}\Phi A_0^aA_0^a \nonumber\\
& &\hspace*{1.0cm} +\frac{1}{2} (D_iA_0^a)^2+\frac{1}{2}m_D^2A_0^aA_0^a+
\frac{1}{4}\lambda_A(A_0^aA_0^a)^2 \,\,\biggr] ,
\la{3dtheory}
\ea
where $F^a_{ij}=\del_iA_j^a-\del_jA_i^a+g_3\epsilon^{abc}A^b_iA^c_j$,
$D_i\Phi=(\del_i-ig_3\tau^aA^a_i/2)\Phi$, and
$D_iA_0^a=\del_iA_0^a+g_3\epsilon^{abc}A_i^bA_0^c$.
The $\tau^a$:s are the Pauli matrices.
All the fields have the dimension
GeV$^{1/2}$, and $\lambda_3$, $g_3^2$,
$h_3$ and $\lambda_A$ have the dimension GeV.

The theory in eq.~\ref{3dtheory} is a simplified theory in the sense
that the U(1)-subgroup has been neglected. All the
fermions can be included, but they appear only in the parameters
of the effective theory. The relations of the parameters of the
3d theory to those of the Standard Model are given in
refs.~\cite{FKRS1,KRS}, but we do not presently need
these relations. In some cases, the 3d theory could be further
simplified by integrating out also the $A_0^a$-field~\cite{FKRS1,JaKaPa},
resulting in a 3d SU(2)-Higgs model. For generality, we keep
the $A_0^a$-field in the action. In the following,
we calculate the effective potential in the theory defined
by eq.~\ref{3dtheory}. Due to the change in the dimension
of the space in the reduction step, the value of this effective
potential at the minimum is then minus the pressure divided
by the temperature.

The Lagrangian in eq.~\ref{3dtheory} is gauge-invariant in
$\tau$-independent gauge transformations, and gauge fixing
and compensating terms have to be added for perturbative
calculations. We choose these as
\beq
S_\xi=\int\! d^3x\biggl[\frac{1}{2\xi}(\del_iA_i^a)^2+
\del_i\bar{c}^a\del_ic^a+g_3\epsilon^{abc}\del_i\bar{c}^aA_i^bc^c
\biggr] \,\, .
\la{gf}
\eeq
The gauge parameter $\xi$ is a renormalized version of the
corresponding 4d gauge fixing parameter. The covariant gauge
condition of eq.~\ref{gf} has both advantages and disadvantages
compared with other usual gauges, like the $R_\xi$-gauge.
On one hand, it is clearly an allowed choice
of gauge, whereas with the $R_\xi$-gauge,
one is choosing a different gauge for each different
value of $\varphi$ in the effective potential,
which might not be allowed. On the other hand,
the covariant gauge generates extra IR-divergences,
as is seen below.

To calculate the effective potential $V(\varphi)$, one writes
$\Phi=[\phi_3+i\phi_4, \varphi+\phi_1+i\phi_2]^T/\sqrt{2}$
in the action $S+S_\xi$ and neglects terms
linear in quantum fields~\cite{Ja}.
This defines a new theory with the masses
$m_1^2\equiv m_3^2+3\lambda_3\varphi^2$,
$m_2^2\equiv m_3^2+\lambda_3\varphi^2$,
$m_L^2\equiv m_D^2+h_3\varphi^2$, and
$m_T^2\equiv g_3^2\varphi^2/4$.
The propagators are
\ba
\contract{{\overline{c}^a(k)}{c^b}}(k) & = &
\frac{\delta^{ab}}{k^2}\nonumber \\
\contract{{\phi_1(-k)}{\phi_1}}(k) & = &
\frac{1}{k^2+m_1^2}\nonumber \\
\contract{{A_0^a(-k)}{A_0^b}}(k) & = &
\frac{\delta^{ab}}{k^2+m_L^2}\nonumber \\
\contract{{A_i^a(-k)}{A_j^b}}(k) & = & \delta^{ab}\left[
\frac{\delta_{ij}-k_ik_j/k^2}{k^2+m_T^2}+
\xi\frac{k_ik_j}{k^2}\frac{k^2+m_2^2}{k^2(k^2+m_2^2)+
\xi m_T^2m_2^2}\right] \la{propagators} \\
\contract{{\phi_G(-k)}{\phi_G}}(k) & = & \frac{k^2+\xi m_T^2}{k^2(k^2+m_2^2)+
\xi m_T^2m_2^2}\,\, , \,\,\,\,\,\,\,\, G=2,3,4 \nonumber \\
\contract{{\phi_2(k)}{A^3_i}}(-k) & = &
-\contract{{\phi_3(k)}{A^2_i}}(-k) =
-\contract{{\phi_4(k)}{A^1_i}}(-k) = \frac{i\xi m_Tk_i}{k^2(k^2+m_2^2)+
\xi m_T^2m_2^2} \,\, .\nonumber
\ea
If $m_2^2$ is negative, a small imaginary part has to be added
to it to define the loop integrals. One then calculates
all the one-particle-irreducible
vacuum diagrams of the theory to a desired order
in the loop expansion. Non-vanishing 2-loop contributions
arise from the diagrams of Fig.~1. The method of calculation is to write
\beq
\frac{1}{k^2(k^2+m_2^2)+\xi m_T^2m_2^2}=
\frac{1}{m_2^2(R_+^2-R_-^2)}\left[\frac{1}{k^2+m_2^2R_-^2}-
\frac{1}{k^2+m_2^2R_+^2}\right]
\,\, , \la{method}
\eeq
where $R^2_{\pm}=1/2\pm\sqrt{1/4-\xi(m_T/m_2)^2}$,
and to use standard Landau-gauge values of integrals~\cite{FKRS1},
with dimensional regularization. The 2-loop effective
potential is presented in the Appendix.




Next, the effective potential has to be renormalized.
Many of the individual 2-loop diagrams contributing to $V(\varphi)$
include gauge-dependent divergent pieces, but all these cancel,
leaving the gauge-independent divergence
\beq
\frac{\hbar^2}{16\pi^2}\frac{\mu^{-4\epsilon}}{4\epsilon}\biggl\{
\frac{\varphi^2}{2}\biggl[
\frac{39}{16}g_3^4+9\lambda_3 g_3^2-12\lambda_3^2+
12h_3g_3^2-6h_3^2\biggr]+
3g_3^2m_3^2+6g_3^2m_D^2\biggr\}
\,\, . \la{counterterm}
\eeq
The piece multiplying $\varphi^2$  is removed by mass renormalization,
and the coupling constants are RG-invariant. Note that if the vacuum terms
are renormalized by just removing the $1/\epsilon$-pieces, the value
of the effective potential at the minimum becomes $\mu$-dependent.
The $\varphi$-dependent part of $V(\varphi)$ is $\mu$-independent,
since the $\mu$-dependence of the renormalized mass squared
$m_3^2(\mu)$ cancels the $\log(\mu)$-terms of the 2-loop graphs.
The renormalized effective potential is obtained from the
expressions in the Appendix
by replacing $m_3^2$ with $m_3^2(\mu)$, and
by ignoring the $1/\epsilon$-piece in the function $H(m_a,m_b,m_c)$.

\section{Equation of state}
\la{eos}

To calculate the value of $V(\varphi)$ at the minimum
in powers of $\hbar$,
we write $V(\varphi)$ as $V=V_0+\hbar V_1+\hbar^2 V_2$,
with the classical part $V_0=m_3^2\varphi^2/2+\lambda_3\varphi^4/4$.
The location of the minimum is determined from
\beq
0=\frac{d}{d\varphi}\left[\left.V_0(\varphi)\right|_{
\varphi=\varphi_0+\hbar\varphi_1+\hbar^2\varphi_2}+
\hbar \left.V_1(\varphi)\right|_{\varphi=\varphi_0+\hbar\varphi_1}+
\hbar^2 \left.V_2(\varphi)\right|_{\varphi=\varphi_0}\right]
\,\, .
\la{minimum}
\eeq
The $\hbar^0$-term of this equation reads
$V_0'(\varphi_0)=\varphi_0m_2^2=0$, yielding
the two solutions $\varphi_0=0$ and $\varphi_0^2=-m_3^2/\lambda_3$.
Let us assume that $m_3^2<0$, and inspect first the broken minimum.
{}From eq.~\ref{minimum}, the corrections to $\varphi_0$ are
\beq
\varphi_1=-\frac{V'_1}{V''_0}\hspace*{1.4cm};\hspace*{1.4cm}
\varphi_2=
-\frac{1}{V''_0}\frac{d}{d\varphi}
\left[V_2-\frac{1}{2}\frac{(V'_1)^2}{V''_0}\right]
\,\, ,
\la{loci}
\eeq
where each expression is evaluated at $\varphi_0$.
The value of the 3d effective potential at the minimum is then
\beq
-\frac{p(T)}{T}=V_0(\varphi_0)+\hbar\biggl[
V_1+\varphi_1V'_0\biggr]_{\varphi=\varphi_0}+
\hbar^2\biggl[V_2-\frac{1}{2}\frac{(V'_1)^2}{V''_0}
+\varphi_2V'_0\biggr]_{\varphi=\varphi_0}
\,\, . \la{value}
\eeq
This should be gauge-independent.

At $\varphi_0$, $V'_0=0$ by definition, so that
the terms multiplying $\varphi_1$ and $\varphi_2$ in
eq.~\ref{value} should be put to zero. However,
these terms are kept for the moment, for the following reason.
In the limit $m_2^2\to 0$ the Goldstone mode propagator is
\beq
\frac{1}{k^2}+\frac{\xi m_T^2}{k^4}
\,\, .
\eeq
Because this is IR-divergent inside loop integrals, the
2-loop effective potential diverges in the limit $m_2^2\to 0$.
The divergences from the individual diagrams which are of the
form $\xi^{5/2}m_2^{-1}$, $\xi^{7/4}m_2^{-1/2}$, or $\xi^2\log(m_2)$,
cancel, but the divergent piece
\beq
\frac{1}{16\pi^2}\biggl[\frac{9}{8}
\lambda_3 m_T^2\xi^{3/2}\biggl(\frac{m_T}{m_2}\biggr)-
\frac{9}{2\sqrt{2}}\lambda_3 m_T\xi^{3/4}\biggl(\frac{m_T}{m_2}\biggr)^{1/2}
\biggl(m_1+m_T\frac{g_3^2}{2\lambda_3}+m_L\frac{h_3}{\lambda_3}\biggr)\biggr]
\la{diver}
\eeq
remains. These problems show
up even at the 1-loop level: the $\xi$-dependent
part of~$V_1$ is finite but non-analytic,
\beq
V_1^{(\xi)}\propto-\xi^{3/4}m_2^{3/2} \,\, .
\la{div1}
\eeq
Due to the equation ${d}m_2^r/{d\varphi}=rm_2^{r-2}\lambda_3\varphi$,
it is seen from eqs.~\ref{loci},~\ref{diver} and~\ref{div1}
that the correction~$\varphi_1$
diverges as $-\xi^{3/4}m_2^{-1/2}$, and
$\varphi_2$ might diverge as $\xi^{3/2}m_2^{-3}$.
However, it can easily be seen from eq.~\ref{1loop} that the term
\beq
-\frac{1}{2}\frac{(V'_1)^2}{V''_0}
\la{corr}
\eeq
exactly cancels the divergent piece of $V_2$, shown
in eq.~\ref{diver}. Consequently,
it turns out that~$\varphi_2$ only diverges as
$-\xi m_2^{-1}$, so that being multiplied
by $V'_0\propto m_2^2$ in eq.~\ref{value}, it does not
contribute to the value of the effective potential at
the minimum, where $m_2^2=0$. In other words, the terms
$\varphi_1V'_0$ and $\varphi_2V'_0$ can safely be put to zero,
when the IR-divergences are ``regularized'' by handling
the classical minimum as the limit $m_2\to 0^+$.
Let us mention that in the Landau-gauge ($\xi =0$),
$\varphi_1$ and $\varphi_2$ are finite.

In addition to cancelling all the divergences,
the term in eq.~\ref{corr} also cancels the finite gauge-dependent
piece of $V_2$ in eq.~\ref{value}. As mentioned,
the gauge dependence of $V_1$ is of the form $\xi^{3/4}m_2^{3/2}$
near $\varphi_0$, so that this vanishes, too. Thus,
we have verified explicitly that the pressure at the broken minimum
is gauge-independent to order $\hbar^2$.

At the symmetric minimum $\varphi=0$, to 2-loop order
there are no IR-divergences in~$V_2$,
and the effective potential is gauge-independent in itself.
The ``correction term'' in eq.~\ref{corr}  vanishes.
If $m_3^2<0$, the effective potential becomes complex,
in accordance with the fact that the symmetric minimum is,
at the tree level, unstable for $m_3^2<0$.

\medskip

In addition to $\xi$, all the physical
quantities should be independent of the renormalization
scale $\mu$. As mentioned after eq.~\ref{counterterm},
in dimensional regularization the dimensionally
reduced effective theory produces a $\mu$-dependent unphysical vacuum
term to the 3d effective potential,
corresponding to a zero-temperature entropy.
This term is actually removed by a counterterm produced by the
dimensional reduction step, but on the other hand there
is an undetermined vacuum energy density in the full 4d theory.
Apart from the vacuum term, eq.~\ref{value} is $\mu$-independent
to order $\hbar^2$, since the $\varphi$-dependent
part of $V(\varphi)$ is so.
In particular, the difference between the pressures
of the symmetric and broken phases, determining
the critical temperature and the latent heat of
the phase transition, is independent of both $\mu$
and $\xi$ to order $\hbar^2$, as it should.

\section{Gauge-independent effective potential}
\la{giep}

Let us calculate the gauge-independent generating functional defined
in eq.~\ref{desy}, and the value of the corresponding effective
potential at the minimum, in powers of $\hbar$
inside the 3d theory. Since we are only interested
in the effective potential, $J$ may be chosen as a constant.
Defining $w(J)\equiv\widetilde{W}[J]/V$ and
$dw/dJ=\sigma$, the value of the effective potential
$\widetilde{V}(\sigma)=w-\sigma J$ at the minimum is formally
\beq
\left.\widetilde{V}(\sigma)\right|_{d\widetilde{V}/d\sigma=0}=
w(0) \,\,  . \la{formal}
\eeq
To see how this equation arises in powers of $\hbar$, write
$w(J)=w_0+\hbar w_1+\hbar^2 w_2$. For a Legendre transformation,
the source $J=J_0+\hbar J_1+\hbar^2 J_2$ has to be solved from the
equation $dw/dJ=\sigma$. With $J_0$ determined from $w'_0(J_0)=\sigma$,
the effective potential is then
\beq
\widetilde{V}(\sigma)=
w_0(J_0)-\sigma J_0+\hbar w_1(J_0)+
\hbar^2\biggl[w_2-\frac{1}{2}\frac{(w'_1)^2}{w''_0}\biggr]_{J=J_0}
\,\, .
\eeq
Calculating the value of $\widetilde{V}(\sigma)$ at the minimum
as in eqs.~\ref{minimum}-\ref{value}, finally gives the result
$\widetilde{V}(\mbox{min})=w_0(0)+\hbar w_1(0)+\hbar^2 w_2(0)$
as in eq.~\ref{formal}.
Below, we see explicitly that $w_0(0)+\hbar w_1(0)+\hbar^2 w_2(0)$
is given by eq.~\ref{value},
so that as far as the equation of state is concerned, the conventional
effective potential $V(\varphi)$ and the gauge-independent
effective potential $\widetilde{V}(\sigma)$
should lead to the same physics.

To calculate $w(J)$ [and, in particular, $w(0)$],
we need to perform the path integral
\beq
\int{\cal D}\phi{\cal D}A\exp( -S[\phi,A])
\la{integral}
\eeq
using the loop expansion, and then
to substitute $m_3^2\to m_3^2+2J$.
We first discuss the broken minimum.
Let us make the change of variables $\Phi=(0,\varphi_0)^T/\sqrt{2}+\Phi'$
in eq.~\ref{integral}, where $\varphi_0=\sqrt{-m_3^2/\lambda_3}$
is the location of the classical broken minimum.
Since this is a stationary point of the Lagrangian,
all terms linear in quantum fields disappear from
the action (apart from counterterms contributing to higher order
than $\hbar^2$). This means that we have exactly the same Lagrangian
as in the case of calculating the effective potential $V(\varphi)$ at
$\varphi=\varphi_0$. However, the set of diagrams is different, since
in the calculation of the effective potential, only
one-particle-irreducible graphs are included~\cite{Ja}.
Now we have to include all the
connected graphs, which means that one-particle-irreducible
graphs have to be supplemented by graphs of the type
shown in Fig.~2. These graphs can be easily
calculated in the same gauge which was used
before. The result is simple: these graphs contribute the
amount $-(V'_1)^2/(2V''_0)$. Hence, $w(0)$
equals the gauge-independent value of eq.~\ref{value}, as promised.




When constructing the gauge-independent effective
potential in the symmetric phase, there is the complication
that the tree-level part of the
generating function $w(J)$ vanishes, $w_0(J)=0$.
Therefore, the calculations after eq.~\ref{formal}
have to be modified. Nevertheless, it is still true
that $w(0)$ as calculated from
the loop expansion equals the value of the
conventional effective potential at $\varphi=0$.

\section{Other physical quantities in powers of $\hbar$}
\la{physical}

Having constructed an expression for the value of the 3d effective
potential at its minima, which is independent of $\xi$ and independent
of $\mu$ to order $\hbar^2$, we could express the 3d parameters
in terms of the 4d parameters and temperature, and study the properties of
the EW phase transition.
Of course, it is expected that this investigation breaks down
at some order, due to the IR-divergences of finite-temperature
field theory. Let us see, how far we can go without problems.
{}From the difference $\Delta p(T)$
of the pressures of the symmetric and broken phases, we
can in principle calculate
the critical temperature~$T_c$ and the latent heat $L$ to order $\hbar^2$.
The former is the solution of the equation $\Delta p(T_c)=0$,
and the latter is given by $L=T_c\Delta p'(T_c)$.
Unfortunately, the zeroth order term $T_0$ of the critical
temperature~$T_c=T_0+\hbar T_1+\hbar^2 T_2$
is given as the solution of the equation
\beq
m_3^2(T_0,\mu)=0 \,\, , \la{T0}
\eeq
and the vanishing of $m_3^2$ leads to singularities at high enough orders.
The first order term~$T_1$ is still finite, and given by
\beq
T_1=-\frac{\Delta p_1(T_0)}{\Delta p_0'(T_0)}=
\frac{3}{4\pi}h_3m_D\left.\left(\frac{dm_3^2}{dT}
\right)^{-1}\right|_{T=T_0} \,\, .
\eeq
To leading order in the coupling constants, this agrees with the
result in ref.~\cite{Ar}. However, the second order term $T_2$,
which includes a finite piece cancelling the $\mu$-dependence of~$T_0$
arising from eq.~\ref{T0}, also includes a logarithmic divergence
$T_2\propto \log (\mu/m_3)$. Hence, $T_c$ is not calculable
to order $\hbar^2$. As to the latent heat, both the zeroth order
term $L_0=T_0\Delta p_0'(T_0)$
and the first order term
\beq
L_1=-\Delta p_1(T_0)+
T_0\left.\left(\Delta p_1'-\frac{\Delta p_1\Delta p_0''}
{\Delta p_0'}\right)\right|_{T=T_0}
\eeq
vanish. For $L_0$ this
is natural, but for $L_1$ it is not,
since the 1-loop effective potential~$V(\varphi)$
in the Landau gauge has a barrier between the symmetric and broken minima,
suggesting a first-order phase transition.
The second order term $L_2$ is finite~(after
cancellations of various divergences, proportional
to $m_3^{-2}$, $m_3^{-1}$ and $\log m_3$),
but it includes an imaginary part:
\ba
L_2 & = & \frac{T_0^2}{16\pi^2}\left\{
-\frac{\lambda_3}{2}\left[\frac{4}{3}+i\left(
\frac{g_3^3}{4\lambda_3^{3/2}}+\frac{2^{3/2}}{3}\right)\right]^2
\right.\nonumber\\
 & & \hspace{1.3cm}\left. \rule{0in}{4.0ex}
+\frac{1}{4\lambda_3}
\left(\frac{51}{16}g_3^4+9\lambda_3 g_3^2-12\lambda_3^2\right)
\right\}\left.\left(\frac{dm_3^2}{dT}\right)
\right|_{T=T_0}
\,\, .
\ea
The real part of $L_2$ behaves qualitatively
like the $g^4$, $\lambda^2$ -curve in Fig.~6 of ref.~\cite{FoHe},
but the numerical value is much smaller.
We conclude that little constructive information concerning
the actual EW phase transition can be obtained from the present approach.

Even if the physical properties of the EW phase transition
cannot be reliably calculated in powers of $\hbar$, it might be hoped
that quantities related solely to the broken phase, away from
the critical temperature, could be calculated.
The problem is that there
may be $\mu$-dependent vacuum parts in these quantities
as in the value of pressure. However,
when one is comparing perturbation theory with lattice calculations,
$\mu$ can be fixed~\cite{FKRS2},
and the vacuum parts are not a problem. To illustrate the
convergence of these calculations, we note that for
$m_H=80$ GeV, $\mu=100$ GeV and $T=60$ GeV,
the $\hbar^0$, $\hbar^1$ and $\hbar^2$ contributions to
$\langle\Phi^{\dagger}\Phi\rangle$ are $[25.47$, $3.76$
and $0.39]\times 10^3$~GeV$^2$ (for these parameters,
$T_c\approx 170$ GeV). These values seem to
indicate reasonable convergence. In the next Section, the
convergence of perturbation theory in the broken phase
is studied from a different point of view.

\section{Convergence in the broken phase}
\la{con}

Above, we were careful to calculate quantities which
are gauge-independent to
each order in $\hbar$, and we could just hope that this
expansion converges. One could also take a different approach,
trying to extremize convergence but not being too careful
about the gauge dependence. The possible gauge dependence
of the physical observables obtained in this way could then
be used as a measure of the convergence of the expansion.

Good convergence of perturbation theory requires that
the coupling constants, and the higher order contributions
including logarithms of the renormalization scale~$\mu$, are small.
The latter requirement can be satisfied by making
a renormalization group~(RG) improvement to the naive calculation.
In practice, this can be implemented by
choosing $\mu$ to correspond to
a typical mass scale appearing in the
propagators. When calculating the effective potential in a large
range of $\varphi$,
the masses in the propagators depend on the shifted
field $\varphi$, and hence $\mu$ should also be chosen to depend
on $\varphi$. When this is consistently implemented,
one arrives  at the RG-improved
effective potential~\cite{FKRS1,DRTJ}.

In ref.~\cite{La}, the physical observable
$dp/dm^2=-\langle\Phi^{\dagger}\Phi\rangle$ was chosen
as the indicator of the convergence
of the RG-improved loop expansion. The value of
$\langle\Phi^{\dagger}\Phi\rangle$ was calculated
numerically from the RG-improved 2-loop effective potential.
Due to RG-improvement, and the fact that
one is calculating the location of the broken minimum exactly
instead of using eq.~\ref{loci}, the result includes
a certain subset of contributions from higher powers of $\hbar$.
Since $\langle\Phi^{\dagger}\Phi\rangle$ is gauge-independent
in the full theory, the gauge dependence of the mixed-order
result tells something about the order of magnitude of the
missing terms. It is seen in Fig.~3 of ref.~\cite{La} that
$\sqrt{2\langle\Phi^{\dagger}\Phi\rangle}$
depends much less on the gauge fixing parameter than,
for instance, the location of the broken minimum,
which is not gauge-independent.
To give a numerical illustration,
for $\xi=0$ and the parameters cited at the
end of Sec.~\oref{physical}, the
value of $\langle\Phi^{\dagger}\Phi\rangle$ is
$29.64\times 10^3$~GeV$^2$, and the uncertainty due to
gauge dependence is of order 1 percent. This gives support to the
argument that the RG-improved loop expansion converges well
deep in the broken phase.

\section{Conclusions}
\la{clu}

By calculating consistently in powers of $\hbar$, we have been
able to derive a gauge-independent expression for the pressure
of EW matter at high temperature. Unfortunately, the physical
properties of the EW phase transition cannot be reliably calculated.
However, quantities related solely to the broken phase are
calculable in powers of $\hbar$,
and the convergence of perturbation theory should be
reasonably good.

\section*{Acknowledgements}
I am grateful to K.~Kajantie and M.~Shaposhnikov
for discussions.

\newpage

\setlength{\parindent}{0mm}
{\Large{\bf Appendix}}
\vspace*{2mm}

We present here the 2-loop effective potential
of the theory defined by eqs.~\ref{3dtheory} and~\ref{gf}.
The tree-level part is
\beq
V_0(\varphi)=\frac{1}{2}m_3^2\varphi^2+\frac{1}{4}\lambda_3\varphi^4
\,\, .
\eeq
With the functions $R_{\pm}$ defined in eq.~\ref{method},
the 1-loop contribution to the effective potential is
\beq
V_1(\varphi)=
-\frac{1}{12\pi}\bigl[6m_T^3+3m_L^3+m_1^3+3m_2^3(R_+^3+R_-^3)\bigr]
\,\, . \la{1loop}
\eeq
In dimensional regularization, there are no divergences
in $V_1(\varphi)$. To present the 2-loop contribution, we
use the function
\beq
H(m_a,m_b,m_c)=\frac{1}{16\pi^2}\Biggl[\frac{1}{4\epsilon}+
\log\biggl(\frac{\mu}{m_a+m_b+m_c}\biggr)+\frac{1}{2}\Biggr]\,\, ,
\la{H}
\eeq
arising from the sunset diagrams in Fig.~1. For
brevity, we also use the functions $I(m_a,m_b,m_c)$,
$D_{VVS}(M,M,m)$ and $D_{SSV}(m_a,m_b,M)$ defined
in ref.~\cite{FKRS1}. These functions are
shorthands for certain combinations of masses and logarithms,
and the function $H(m_a,m_b,m_c)$ appears linearly in them.
Four more functions, denoted by ${\cal A}$, ${\cal B}$,
${\cal C}$ and ${\cal D}$, are defined at the
end of this Appendix. A common factor $1/16\pi^2$,
and the factor $\mu^{-4\epsilon}$ multiplying $H(m_a,m_b,m_c)$,
have been omitted from all the formulas below; accordingly,
the factor $1/16\pi^2$ in eq.~\ref{H} should also be left out.
The divergent part of $V_2(\varphi)$
is obtained from the coefficient of the function $H(m_a,m_b,m_c)$,
and is displayed in eq.~\ref{counterterm}. To get the
renormalized expression for $V_2(\varphi)$,
the term $1/\epsilon$ is to be omitted from eq.~\ref{H},
and $m_3^2$ is to be replaced by $m_3^2(\mu)$. Let us note
that due to the identities $R_{+}^2+R_{-}^2=1$ and
$R_{+}^2R_{-}^2=\xi(m_T/m_2)^2$, it might be possible
to simplify some of the formulas below.
We denote $m_2R_{\pm}$ by $m_{\pm}$. Finally,
for pure SU(2)-Higgs theory without the $A_0^a$-field, the graphs
(g3), (e2), (g4), (b) and (e1) are left out.

\makeatletter
\def\[{\relax\ifmmode\@badmath\else\begin{trivlist}\item[]\leavevmode
 \hbox to\linewidth\bgroup$ \displaystyle
 \hskip\mathindent\bgroup\fi}
\def\]{\relax\ifmmode \egroup $\hfil \egroup \end{trivlist}\else \@badmath \fi}
\def\equation{\refstepcounter{equation}\trivlist \item[]\leavevmode
 \hbox to\linewidth\bgroup $ \displaystyle
\hskip\mathindent}

\def\endequation{$\hfil \displaywidth\linewidth\@eqnnum\egroup \endtrivlist}
\def\eqnarray{\stepcounter{equation}\let\@currentlabel=\theequation
\global\@eqnswtrue
\global\@eqcnt\z@\tabskip\mathindent\let\\=\@eqncr
\abovedisplayskip\topsep\ifvmode\advance\abovedisplayskip\partopsep\fi
\belowdisplayskip\abovedisplayskip
\belowdisplayshortskip\abovedisplayskip
\abovedisplayshortskip\abovedisplayskip
$$\halign to
\linewidth\bgroup\@eqnsel\hskip\@centering$\displaystyle\tabskip\z@
 {##}$&\global\@eqcnt\@ne \hskip 2\arraycolsep \hfil${##}$\hfil
 &\global\@eqcnt\tw@ \hskip 2\arraycolsep $\displaystyle{##}$\hfil
 \tabskip\@centering&\llap{##}\tabskip\z@\cr}
\def\endeqnarray{\@@eqncr\egroup
 \global\advance\c@equation\m@ne$$\global\@ignoretrue }
\newdimen\mathindent
\mathindent = \leftmargini
\makeatother
\mathindent=0.0cm
\newcommand{\vali}{\vspace*{-5mm}}

\bea
\frac{(d2)}{g_3^2} & = &  m_T^2 \left(2+\frac{\xi^{3/2}}
    {R_{+}+R_{-}}\right)^2
\eea
\vali
\bea
\frac{(h4)}{3\lambda_3/4} & =  & m_1^2+2m_1m_2
    \frac{R_{+}^5-R_{-}^5}{R_{+}^2-R_{-}^2}+5
    m_2^2 \left(\frac{R_{+}^5-R_{-}^5}{R_{+}^2-R_{-}^2}\right)^2
\eea
\vali
\bea
\frac{(f2)}{3g_3^2/8}  & =  & m_T \left(m_1+3
    m_2 \frac{R_{+}^5-R_{-}^5}{R_{+}^2-R_{-}^2}\right)
    \left(2 + \frac{\xi^{3/2}}{R_{+}+R_{-}}\right)
\eea
\vali
\bea
\frac{(g3)}{15\lambda_A/4}  & =  & m_L^2
\eea
\vali
\bea
\frac{(e2)}{3g_3^2}  & =  & m_Lm_T
    \left(2 + \frac{\xi^{3/2}}{R_{+}+R_{-}}\right)
\eea
\vali
\bea
\frac{(g4)}{3h_3/2}  & =  & m_L \left(m_1+3
    m_2 \frac{R_{+}^5-R_{-}^5}{R_{+}^2-R_{-}^2}\right)
\eea
\vali
\bea
\frac{(a)}{-3g_3^4\varphi^2/16} & = &
    D_{VVS}(m_T,m_T,m_1) \\
&  & \hspace*{-10mm} +\frac{\xi}{2(m_{+}^2-m_{-}^2)}
    \Biggl\{\frac{R_{+}^2}{R_{-}^2}\Bigl[D_{SSV}(0,m_1,m_T)-
    D_{SSV}(m_{-},m_1,m_T)\Bigr]-({+}\leftrightarrow {-})\Biggr\} \\
&  & \hspace*{-10mm} +\frac{\xi^2}{(m_{+}^2-m_{-}^2)^2}\Biggl\{
    \frac{R_{+}^4}{R_{-}^4}\Bigl[
    I(0,0,m_1)-2 I(0,m_{-},m_1)+
    I(m_{-},m_{-},m_1)\Bigr] \\
&  & \hspace*{-10mm} -\Bigl[ I(0,0,m_1)-I(0,m_{+},m_1)-
I(0,m_{-},m_1)+I(m_{-},
    m_{+},m_1)\Bigr] +({+}\leftrightarrow {-})\Biggr\}
\eea
\vali
\bea
\frac{(b)}{-3 h_3^2 \varphi^2} & = &  H(m_1,m_L,m_L)
\eea
\vali
\bea
\frac{(c)}{-3 \lambda_3^2 \varphi^2} & = &  H(m_1,m_1,m_1) \\
&  & \hspace*{-10mm} +\frac{1}{(R_{+}^2-R_{-}^2)^2}\left[
    R_{+}^8 H(m_{+},m_{+},m_1)-
    R_{-}^4 R_{+}^4 H(m_{-},m_{+},m_1) +
    ({+}\leftrightarrow {-})\right]
\eea
\vali
\bea
\frac{(e1)}{-3g_3^2/2} & = &
    (m_T^2-4m_L^2)H(m_T,m_L,m_L)+2m_Tm_L-m_L^2  \\
&  & \hspace*{-10mm} +\frac{4\xi}{m_{+}^2-m_{-}^2}
    \Biggl\{\frac{R_{+}^2}{R_{-}^2}\Bigl[I(0,m_L,m_L)-
    I(m_{-},m_L,m_L)\Bigr]-({+}\leftrightarrow {-})\Biggr\} \\
&  & \hspace*{-10mm} -\xi m_L^2
    +\frac{\xi^2m_T^2}{R_{+}^2-R_{-}^2}\Bigl[
    H(m_-,m_L,m_L)-H(m_+,m_L,m_L)\Bigr]
\eea
\vali
\bea
\frac{(f1)}{-3g_3^2/8} & = &
    {\cal A}(m_1,m_2,m_T)+\frac{1}{R_{+}^2-R_{-}^2}\Bigl[R_{+}^4
    {\cal A}(m_{+},m_2,m_T)
    -R_{-}^4{\cal A}(m_{-},m_2,m_T)\Bigr]
\eea
\vali
\bea
\frac{(gh)}{3g_3^2m_T^2/4} & = &
    H(m_T,0,0)+\frac{\xi^2}{R_{+}^2-R_{-}^2}\Bigl[
    H(m_{-},0,0)-
    H(m_{+},0,0)\Bigr]
\eea
\vali
\bea
\frac{(x1)}{3 \lambda_3\xi^2 m_T^4/(m_{+}^2-
    m_{-}^2)^2} & = &  (m_{+}-m_{-})^2  \\
& & \hspace*{-10mm} +\left[(2 m_{-}^2-
    m_1^2) H(m_1,m_{-},m_{-}) -(m_2^2-m_1^2)H(m_1,m_{-},m_{+}) +
    ({+}\leftrightarrow {-})\right]
\eea
\vali
\newcommand{\sama}{(+\leftrightarrow -)}
\bea
\frac{(x2)}{-6 \lambda_3 \xi m_T^2m_2^2/(m_{+}^2-
    m_{-}^2)^2} & = &  -R_{-}^4
    \biggl\{(m_{-}-m_{+})(m_{-}-m_1) \\
& & \hspace*{-10mm} +(m_{-}^2-m_1^2)\Bigl[H(m_1,m_{-},m_{-})
     -H(m_1,m_{+},m_{-})\Bigr]\biggr\} + \sama
\eea
\vali
\bea
\frac{(x3)}{3 g_3^2\xi m_T^2/
    4(m_{+}^2-
    m_{-}^2)} & = &  D_{SSV}(m_1,m_{-},m_T)-
    D_{SSV}(m_1,m_{+},m_T)
    +\left\{\frac{\xi R_{+}^2}{R_{+}^2-
    R_{-}^2}\Biggl[\right.\\
& & \hspace*{-40mm}\left.(m_{-}-m_{+})(m_{-}-m_1)
    +(2 m_{-}^2-m_1^2) H(m_1,m_{-},m_{-})
    -(m_2^2-m_1^2)H(m_1,m_{+},m_{-})\right.\\
& & \hspace*{-40mm}+\left.\frac{4}{m_{-}^2}\Bigl[I(m_{-},0,m_1)-
    I(m_{-},m_{-},m_1)-I(m_{+},
    0,m_1)+I(m_{+},m_{-},m_1)\Bigr]\Biggr]-\sama\right\}
\eea
\vali
\bea
\frac{(x4)}{-3g_3^2 \xi^2 m_T^2/8} & = &
    {\cal B}(m_1,m_2,m_T)-\frac{2}{R_{+}^2-R_{-}^2}\left[R_{+}^4
    {\cal B}(m_{+},m_2,m_T)-
    R_{-}^4 {\cal B}(m_{-},m_2,m_T)\right]
\eea
\vspace*{-3mm}
\bea
\frac{(x5)}{3g_3^2\xi^2m_T^2/(m_{+}^2-
    m_{-}^2)^2} & = &  m_{-}^4+m_{+}^4-m_{-}m_{+}m_2^2
    +\biggl\{I(m_{-},m_T,m_{-}) \\
& & \hspace*{-40mm}  -I(m_{-},m_T,m_{+})+m_T^2 m_{-}^2
    \Bigl[H(m_{-},m_{+},m_T)-H(m_{-},m_{-},m_T)\Bigr]+\sama \biggr\}
\eea
\vali
\bea
\frac{(d1)}{g_3^2} & = & \frac{m_T^2}{8}\Bigl[63H(m_T,m_T,m_T)
    -3H(m_T,0,0)-41\Bigr]\\
& & \hspace*{-1mm} -\frac{3\xi}{2(R_{+}^2-R_{-}^2)}\left[R_{+}^2
    {\cal C}(m_T,m_{-})-R_{-}^2
    {\cal C}(m_T,m_{+})\right] \\
& & \hspace*{-1mm} + \frac{3\xi^2}{2(R_{+}^2-R_{-}^2)^2}
    \left[R_{+}^4{\cal D}(m_T,m_{-},
    m_{-})-R_{+}^2 R_{-}^2
    {\cal D}(m_T,m_{-},m_{+})+\sama\right]
\eea
\vspace*{2mm}

\noindent The functions ${\cal A}$, ${\cal B}$, ${\cal C}$ and  ${\cal D}$
appearing above are
\bea
{\cal A}(m_1,m_2,m_T) & = & \frac{1}{R_{+}^2-R_{-}^2}
    \left[R_{+}^4 D_{SSV}(m_1,m_{+},
    m_T)-R_{-}^4 D_{SSV}(m_1,m_{-},m_T)\right]\\
& & \hspace*{-35mm} +\frac{\xi}{(R_{+}^2-R_{-}^2)^2}\biggl\{-R_{+}^2 R_{-}^4
    \Bigl[2 m_{-}^2+
    m_1 m_{-}-D_{SSV}(m_1,m_{-},m_{-})-
    (m_{-}^2+2 m_1^2) H(m_1,m_{-},m_{-})\Bigr]\\
& & \hspace*{-35mm} +R_{+}^6 \Bigl[2 m_{-}( m_{+}+m_1)-m_1 m_{+}-
    D_{SSV}(m_1,m_{+},m_{-}) \\
& & \hspace*{-35mm}  +(m_{-}^2-2
    m_{+}^2-2 m_1^2) H(m_1,m_{-},m_{+})\Bigr]+\sama\biggr\}
\eea
\bea
{\cal B}(m_1,m_2,m_T)& = &\frac{1}{(m_{+}^2-m_{-}^2)^2}
    \biggl\{-m_1^2 (m_{+}-m_{-})^2\\
& & \hspace*{-20mm} +\Bigl[(m_1^2-m_{-}^2)^2 H(m_1,m_{-},m_{-})-
    (m_1^2-m_{-}^2) (m_1^2-m_{+}^2)
    H(m_1,m_{-},m_{+})+\sama\Bigr]\biggr\}
\eea
\bea
{\cal C}(m_T,m) & = &\frac{1}{2 m^2}\left[-m_T^4 H(m_T,0,0)+
    (m_T^2-m^2)^2 H(m_T,m,0)+
    m m_T \left(m_T^2+\frac{19m^2}{3}\right)\right]
\eea
\bea
{\cal D}(m_T,m_1,m_2) & = & \frac{m_T^2}{4 m_1^2 m_2^2}\biggl\{
    (m_T^2-m_1^2)^2 H(m_T,m_1,0)+
    (m_T^2-m_2^2)^2 H(m_T,m_2,0)\\
& & \hspace*{-20mm} - m_T^4 H(m_T,0,0)-\left[m_T^2-(m_1-m_2)^2\right]
    \left[m_T^2- (m_1+m_2)^2\right]H(m_T,m_1,m_2)\biggr\}\\
& & \hspace*{-20mm} +\frac{1}{4 m_1 m_2}\left[{m_T^3(m_T+ m_1+m_2)-
    m_T^2 (m_1^2+m_2^2)-8 m_1^2 m_2^2/3}\right]\,\, .
\eea
\newpage

\end{document}